\documentstyle[sprocl,psfig]{article}
\def\NPB{{\em Nucl. Phys.} B}
\def\PLB{{\em Phys. Lett.}  B}
\def\PRL{\em Phys. Rev. Lett.} 
\def\PRD{{\em Phys. Rev.} D}
\def\ZPC{{\em Z. Phys.} C}
\def\be{\begin{equation}}
\def\ee{\end{equation}}
\def\bea{\begin{eqnarray}}
\def\eea{\end{eqnarray}}
\def\simlt{\stackrel{<}{{}_\sim}}
\def\simgt{\stackrel{>}{{}_\sim}}
\def\be{\begin{equation}}
\def\ee{\end{equation}}
\def\bear{\be\begin{array}}
\def\eear{\end{array}\ee}
\def\bea{\begin{eqnarray}}
\def\eea{\end{eqnarray}}
\begin{document}
%%%%%%%%%%%%%%%%%%%%%%%%%%% subequations.sty %%%%%%%%%%%%%%%%%%%%%%%%
\catcode`@=11
\newtoks\@stequation
\def\subequations{\refstepcounter{equation}%
\edef\@savedequation{\the\c@equation}%
  \@stequation=\expandafter{\theequation}%   %only want \theequation
  \edef\@savedtheequation{\the\@stequation}% % expanded once
  \edef\oldtheequation{\theequation}%
  \setcounter{equation}{0}%
  \def\theequation{\oldtheequation\alph{equation}}}
\def\endsubequations{\setcounter{equation}{\@savedequation}%
  \@stequation=\expandafter{\@savedtheequation}%
  \edef\theequation{\the\@stequation}\global\@ignoretrue

\noindent}
\catcode`@=12
%%%%%%%%%%%%%%%%%%%%%%%%%%%%%%%%%%%%%%%%%%%%%%%%%%%%%%%%%%%%%%%%%%%%%
\title{CHARGE AND COLOR BREAKING}
\author{J. Alberto CASAS}
\address{Instituto de Estructura de la Materia, Serrano 123 
(CSIC),\\ 28006 Madrid, Spain}
%%%%%%%%%%%%%%%%%%%%%%%%%%%%%%%%%%%%%%%%%%%%%%%%%%%%%%%%%%%%%%
% You may repeat \author \address as often as necessary      %
%%%%%%%%%%%%%%%%%%%%%%%%%%%%%%%%%%%%%%%%%%%%%%%%%%%%%%%%%%%%%%
\maketitle\abstracts{
The presence of scalar fields with
color and electric charge in supersymmetric theories 
makes feasible the existence of dangerous charge and color breaking (CCB)
minima and unbounded from below directions (UFB) in the effective
potential, which would make the standard vacuum unstable. 
The avoidance of these occurrences imposes severe constraints
on the supersymmetric parameter space. We give here a comprehensive
and updated account of this topic.}

\vskip-7.5cm
\rightline{}
\rightline{ IEM-FT-161/97}
\rightline{ July 1997}
\rightline{arch-ive/9707475}

\vspace{5.3cm}

\section{Introduction}
%\subsection{Producing the Hard Copy}\label{subsec:prod}

Experimental observation tells us that color and electric charge are
gauge quantum numbers preserved in nature. From the thoretical point of view, 
in the Standard Model they are certainly conserved in an automatical way
since the only fundamental scalar field is the Higgs boson, a colorless
electroweak doublet. The Higgs potential has a continuum of degenerate
minima, but these are all physically equivalent and one can always
define the unbroken $U(1)$ generator to be the electric charge.
In supersymmetric (SUSY) extensions of the Standard Model things become more 
complicated. First, the Higgs sector must contain for consistency at
least two Higgs doublets $H_1$, $H_2$ (plus perhaps some singlets or
triplets). Hence, one has to check that the minimum of the Higgs potential
$V(H_1,H_2)$ still occurs for values of $H_1$, $H_2$ which are 
apropriately aligned in order to preserve the electric charge; otherwise
the whole electroweak symmetry becomes spontaneously broken.
Second, the supersymmetric theory has a large number of additional charged
and colored scalar fields, namely all the sleptons and squarks,
say $\tilde l_i$, $\tilde q_i$. Consequently one has to verify
that the minimum of the whole potential $V(H_1,H_2, \tilde q_i, 
\tilde l_i)$ still occurs at a point in the field space,
which we will call ``realistic minimum'' in what follows,
where $\tilde q_i,\tilde l_i =0$, thus preserving 
color and electric charge.

The generic situation is that the scalar potential does not present 
just a single minimum,
and, besides the realistic minimum, there is a number of additional
charge and color breaking (CCB) minima.
Then, a reasonable requirement is that the realistic minimum is the
deepest one, i.e the global minimum of the theory.
%
%the standard vacuum corresponds
%to the deepest minimum of $V$, 
This is certainly the usual constraint imposed in the literature and 
represents the most conservative attitude in order to be safe. 
Nevertheless, a situation with CCB minima deeper than the realistic 
minimum could still be acceptable if the cosmology leads the universe
to the latter and this is stable enough. This issue will be discussed
in the last section of this chapter.

CCB minima are not the only disease that the supersymmetric scalar
potential can present. It may also happen that the field space contains
directions along which the potential becomes unbounded form below
(UFB), which is obviously undesirable. Both issues, CCB and UFB, are
closely related, as we will see throughout the chapter.

\vspace{0.2cm}
\noindent
In order to introduce some notation and to illustrate some relevant
aspects and warnings concerning CCB, let us briefly review the 
CCB condition which has been most extensively used in the literature,
namely the ``traditional'' bound, first studied by Frere et al. 
and subsequently by others \cite{Frere,Claudson}.
The {\em tree-level} scalar potential, $V_0$, in the minimal
supersymmetric standard model (MSSM) is given by
\be
\label{V0}
V_0 = V_F + V_D + V_{\rm soft}\;\; ,
\ee
with
\subequations{
\be
\label{VF}
V_F = \sum_\alpha \left| \frac{\partial W}{\partial \phi_\alpha}
\right| ^2\;\;,
\ee
\be
\label{VD}
V_D = \frac{1}{2}\sum_a g_a^2\left(\sum_\alpha\phi_\alpha^\dagger
T^a \phi_\alpha\right)^2\;\; ,
\ee
\bea
\label{Vsoft}
V_{\rm soft}&=&\sum_\alpha m_{\phi_\alpha}^2
|\phi_\alpha|^2\ +\ \sum_{i\equiv generations}\left\{
A_{u_i}\lambda_{u_i}Q_i H_2 u_i + A_{d_i}\lambda_{d_i} Q_i H_1 d_i
\right.
\nonumber \\
&+& \left. A_{e_i}\lambda_{e_i}L_i H_1 {e_i} + {\rm h.c.} \right\}
+ \left( B\mu H_1 H_2 + {\rm h.c.}\right)\;\; ,
\eea}
\endsubequations
where $W$ is the MSSM superpotential
\be
\label{W}
W=\sum_{i\equiv generations}\left\{
\lambda_{u_i}Q_i H_2 u_i + \lambda_{d_i}Q_i H_1 d_i
+ \lambda_{e_i} L_i H_1 e_i \right\} +  \mu H_1 H_2\;\; ,
\ee
$\phi_\alpha$ runs over all the scalar components of the chiral
superfields
and $a, i$ are gauge group and generation indices respectively.
$Q_i$ ($L_i$) are
the scalar partners of the quark (lepton) $SU(2)_L$ doublets     and
$u_i,d_i$ ($e_i$) are the scalar partners of the quark (lepton)
$SU(2)_L$ singlets. In our notation $Q_i\equiv (u_L,\ d_L)_i$,
$L_i\equiv (\nu_L,\ e_L)_i$,
$u_i\equiv {u_R}_i$, $d_i\equiv {d_R}_i$, $e_i\equiv {e_R}_i$. Finally,
$H_{1,2}$
are the two SUSY Higgs doublets. The first observation is that the 
previous potential is extremely
involved since it  has a large number of independent fields.
Furthermore, even assuming universality of the soft breaking terms at
the unification scale, $M_X$, it contains a large number of
independent parameters: $m$, $M$, $A$, $B$, $\mu$, i.e. the universal
scalar and gaugino masses, the universal
coefficients of the trilinear and bilinear scalar terms, and
the Higgs mixing mass, respectively. In addition, there are the
gauge ($g$)
and Yukawa ($\lambda$)
couplings which are constrained by the experimental data. Notice that
$M$ does not appear explicitely in $V_0$, but it does through the
renormalization group equations (RGEs) of all the remaining parameters.

The complexity of $V$ has made that until recently only particular 
directions in the field-space have been explored. The best-known
example of this is the ``traditional'' bound, first studied by Frere 
et al. and subsequently by others \cite{Frere,Claudson}. 
These authors considered just the three fields present in a
particular trilinear scalar coupling, e.g. $\lambda_u A_u Q_u H_2 u$,
assuming
equal vacuum expectation values (VEVs) for them:
\be
\label{frerevevs}
|Q_u| = |H_2| = |u| \;\;,
\ee
where only the $u_L$-component of $Q_u$ takes a VEV in order to cancel
the
D--terms. The phases of the three fields are taken in such way that the
trilinear scalar term in the potential gets negative sign.
Then, the potential (\ref{V0}) gets extremely simplified and
it is easy to show that a very deep
CCB minimum appears {\em unless} the famous constraint
\be
\label{frerebound}
|A_u|^2 \leq 3\left( m_{Q_u}^2 + m_{u}^2 + m_2^2\right)
\ee
is satisfied. In the previous equation $m_{Q_u}^2, m_{u}^2, m_2^2$ are
the
mass parameters of $Q_u$, $u$, $H_2$. Notice from eq.(\ref{V0}) that
$m_2^2$ is the sum of the $H_2$ squared soft mass, $m_{H_2}^2$, plus
$\mu^2$. Similar
constraints for the other trilinear terms can straightforwardly be
written.  These ``traditional'' bounds have extensively been used in
the literature. Notice that the trilinear coefficient, $A$,
plays a crucial role for the appearance of a CCB minimum. This is
logical since the scalar trilinear terms are essentially negative
contributions to the scalar potential (they are negative for a
certain combination of the phases of the fields). However, we will 
see in sect.4 that they are irrelevant for UFB directions.

From the previous bound we can extract two important lessons. First,
many ordinary CCB bounds (as the one of eq.(\ref{frerebound})) come 
from the analysis of particular directions in the
field-space, thus corresponding to {\em necessary but not
sufficient} conditions to avoid dangerous CCB minima.
Consequently a complete analysis requires a more exhaustive exploration
of the field space. 
%Such an analysis is carried out in sections 4-6,
%where a set of complete and optimized bounds is obtained, paying
%special attention to the most powerful one.
Second, the bound of eq.(\ref{frerebound}) has been obtained from the analysis
of the tree-level potential $V_0$. Hence, the {\em radiative 
corrections} should be incorporated in some way. With regard to this
point a usual practice has been to consider the tree-level scalar
potential improved by one-loop RGEs, so that all the parameters
appearing in it (see eq.(\ref{V0}))
are running
with the renormalization scale, $Q$. Then it is demanded that the
previous CCB constraints, i.e. eq.(\ref{frerebound}) and others, are 
satisfied at any scale between $M_X$ and $M_Z$. As we will see in sect.2
this procedure is not correct and leads to an overestimate of the
restrictive power of the bounds. Therefore a more careful treatment of
the radiative corrections is necessary when analyzing CCB bounds.
%This important issue is analyzed in section 2.

The chapter is organized as follows. Section 2 is devoted to analyze
and give prescriptions to handle the above-mentioned issue of the
radiative corrections. Section 3 deals with the Higgs part of the 
potential, which is a requirement for subsequent analyses. In sections
4 and 5 a complete analysis of the UFB and UFB directions of the 
MSSM field space is performed, giving a complete set of optimized
bounds. Special attention will be paid to the most powerful one,
the so-called UFB-3 bound.
The effective restrictive power of these bounds is examined in section 6.
Section 7 is devoted to the bounds that CCB pose on flavour mixing
couplings, which turn out to be surprisingly strong. Finally, the 
cosmological considerations are left for section 8.

\section{The role of the radiative corrections}

As has been mentioned in sect.1, in the CCB analysis the
scalar potential is usually considered at tree-level, improved by
one-loop RGEs, so that all the parameters appearing in it 
(see eq.(\ref{V0})) are running with the renormalization scale, $Q$. 
The two questions that arise are:

\begin{itemize}

\item What is the appropriate scale, say $Q=\hat Q$, to evaluate
$V_0$?

\item How important are the radiative corrections that are being ignored?

\end{itemize}

\noindent
These two questions are intimately related. To understand this it is
important to recall that the {\em exact}
effective potential
\be
\label{VV}
V(Q,\lambda_\alpha(Q),\ m_\beta(Q),\phi(Q)) 
\ee
(in short $V(Q,\phi)$), where 
$\lambda_\alpha(Q),m_\beta(Q)$ are running parameters and masses
and $\phi(Q)$ are the generic classical fields, is scale-independent,
i.e.
\be
\label{Qind}
\frac{d V}{dQ}=0\;\;.
\ee
This property allows in principle any choice of $Q$, and in particular
a different one for each value
of the classical fields, i.e. $Q=f(\phi)$. When analyzing CCB bounds,
one is interested in possible CCB minima , so one has to minimize the 
scalar potential. Denoting by
$\langle \phi\rangle$ the VEVs of the $\phi$--fields obtained from
the minimization of $V$, it is clear from (\ref{Qind}) 
that the two following minimization conditions
\be
\label{mincond1}
\frac{\partial V(Q=f(\phi),\phi)}{\partial \phi} = 0
\ee
\be
\label{mincond2}
\left.\frac{\partial V(Q,\phi)}{\partial \phi}\right|_{Q=f(\phi)} = 0
\ee
yield equivalent results for $\langle \phi\rangle$ (for a more detailed
discussion see refs.\cite{ford,CEQR}).

\vspace{0.3cm}
\noindent
The previous results apply exactly {\em only} to the exact effective
potential. In practice, however, we can only know $V$ with a
certain degree of accuracy in a perturbative expansion. In particular,
at one-loop level
\be
\label{V1def}
V_1=V_0(Q,\phi) + \Delta V_1(Q,\phi)
\ee
where $V_0$ is the (one-loop improved) tree-level potential
and $\Delta V_1$ is the one-loop radiative correction to the
effective potential
\be
\label{DeltaV1p}
\Delta V_1={\displaystyle\sum_{\alpha}}{\displaystyle\frac{n_\alpha}{64\pi^2}}
M_\alpha^4\left[\log{\displaystyle\frac{M_\alpha^2}{Q^2}}
-\frac{3}{2}\right]\;\;.
\ee
Here $M_\alpha^2(Q)$ are the improved tree-level
squared mass eigenstates and $n_\alpha=(-1)^{2s_\alpha} (2s_\alpha+1)$, 
where $s_\alpha$ is the spin of the corresponding particle. 
It is important to notice that $M_\alpha^2(Q)$ are in general
field--dependent quantities since they are the eigenvalues of
the $(\partial^2 V_0/\partial \phi_i\partial\phi_j)$ matrix. Hence, the 
values of $M_\alpha^2(Q)$ depend on the values of the fields and thus
on which direction in the field space is being analyzed.
$V_1(Q,\phi)$ does {\em not} obey eq.(\ref{Qind}) for all values of $Q$.
However, in the region of $Q$ of the order
of the most significant masses appearing in (\ref{DeltaV1p}), the logarithms
involved in the radiative corrections, and the
radiative corrections themselves (i.e. $\Delta V_1$), are minimized, 
thus improving the perturbative expansion. So we expect $V$ to be well 
approximated by $V_1$
and it is not surprising that in that region of $Q$,
$V_1$ is approximately scale-independent \cite{Gamberini,CC}, i.e.
eq.(\ref{Qind}) is nearly satisfied. On the other hand, due to the
smallness of $\Delta V_1$, 
$V_1$ and $V_0$ are, in this region, very similar.
Consequently (always in this region
of $Q$) we can safely approximate $V$ by $V_1$ or 
even\footnote{More precisely, for a choice of $Q$ such that
${\partial\Delta V_1}/{\partial\phi} = 0$ the results from
$V_0$ and $V_1$ are the same. In practice this precise condition
is quite involved and such a degree of precision is not necessary.} 
$V_0$, and
minimize by using either eq.(\ref{mincond1}) or
eq.(\ref{mincond2}), although of course
eq.(\ref{mincond2}) is much easier to handle. This statement can
be numerically confirmed, see e.g. refs.\cite{CEQR,CCB}.

% {}From the these arguments it is sufficiently
%good for our calculation to work just with $V_0$ evaluated at 
%a scale $Q=\hat Q$ of the order of the most significant
%$M_\alpha$ mass appearing in (\ref{DeltaV1p}), i.e.
%$V(\phi)\simeq V_0(\phi, \hat Q)$.

In conclusion, the radiative corrections are reasonably well 
incorporated by using  the tree-level potential $V_0(\phi, \hat Q)$, 
where the renormalization scale 
$\hat Q$ is of the order of the most significant mass, normally
$\hat Q\sim \phi$.
The application of these recipes to our task of determining the
CCB minima and extract the corresponding CCB bounds will be
shown in sects.4,5.

\section{The Higgs potential and the realistic minimum}

The Higgs part of the MSSM potential can be extracted 
(at tree level) from eq.(\ref{V0}). It reads
\bea
\label{Vhiggscomp}
\hspace{-0.3cm}V_{\rm Higgs}&=&m_1^2 |H_1|^2 + m_2^2 |H_2|^2 - m_3^2
\left(\epsilon_{ij}H_1^i H_2^j + {\rm h.c.}\right)
- \frac{1}{2}g_2^2\left|\epsilon_{ij}H_1^i H_2^j \right|^2
\nonumber \\
&+& \frac{1}{8}(g_2^2+g'^2)(|H_2|^4+|H_1|^4)
+ \frac{1}{8}(g_2^2-g'^2)|H_2|^2|H_1|^2,
\eea
where $H_1\equiv (H_1^0,\ H_1^-)$,
$H_2\equiv (H_2^+,\ H_2^0)$,
%
%$H_1\equiv\left(
%\begin{array}{c}H_1^0 \\ H_1^- \end{array}\right)$,
%$H_2\equiv\left(
%\begin{array}{c}H_2^+ \\ H_2^0 \end{array}\right)$,
%
$m_1^2\equiv m_{H_1}^2 + \mu^2$, $m_2^2\equiv m_{H_2}^2 + \mu^2$,
$m_3^2\equiv-\mu B$ and $g_2$, $g'$ are the gauge couplings of 
$SU(2)\times SU(1)_Y$. All these parameters are understood to be 
running parameters evaluated at some renormalization scale $Q$.

Our first interest in $V_{\rm Higgs}$ comes from the fact that 
$V_{\rm Higgs}$ depends not only on the neutral components of $H_1$, 
$H_2$, but also on the charged ones, i.e. $H_1^-$, $H_2^+$. Hence, 
one should check that $\langle H_1^-\rangle$, $\langle H_2^+\rangle$ 
remain vanishing when $V_{\rm Higgs}$ is minimized
(one of them, say $\langle H_2^+\rangle$, can always be chosen as 
vanishing through an $SU(2)$ rotation). Fortunately, it 
is easy to show from (\ref{Vhiggscomp}) that the minimum of 
$V_{\rm Higgs}$ always lies at  $H_2^+ = H_1^- = 0$. So the MSSM 
is safe from this point of view. It is worth remarking that non-minimal
supersymmetric extensions of the standard model do not have this 
nice property, at least in such an automatic way. (This is e.g. 
the case of the so-called 
next-to-minimal supersymmetric standard model (NMSSM),  which contains
an extra singlet in the Higgs sector \cite{ellis}.) Therefore we can
set $H_2^+ = H_1^- = 0$ and focuss our attention on the neutral part
of $V_{\rm Higgs}$, which reads
\bea
\label{Vhiggs}
\hspace{-0.1cm} V_{\rm Higgs}=m_1^2 |H_1|^2 + m_2^2 |H_2|^2 - 
2|m_3^2||H_1||H_2|
+ \frac{1}{8}(g'^2+g_2^2)(|H_2|^2-|H_1|^2)^2 .
\eea
Notice that, since we are interested in the minimization of the
potential, we have implicitely chosen in (\ref{Vhiggs}) a phase 
of $H_1$, $H_2$ such that the mixing term $\propto (\epsilon_{ij}
H_1^i H_2^j + {\rm h.c.})$ gets negative. 

The  second aspect of $V_{\rm Higgs}$ which interests us is that 
$V_{\rm Higgs}$
should develope a minimum at
$|H_1^0|=v_1$, $|H_2^0|=v_2$, such that $SU(2)\times U(1)_Y$ is broken
in the correct way, i.e.
$v_1^2+v_2^2=2M_W^2 / g_2^2\simeq (175\ {rm GeV})^2$. 
This is the realistic minimum
that corresponds to the standard vacuum. This requirement fixes one
of the five independent parameters ($m,M,A,B,\mu$) of the MSSM,
say $\mu$, in terms of the others.
Actually, for some choices of the four remaining parameters
($m,M,A,B$), there is no value of $\mu$ capable of producing the
correct electroweak breaking. Therefore, this requirement restricts
the parameter space further, as is illustrated in Fig.1 (darked
region) with a representative example (which will be discussed in 
detail in sect.6). In addition, the actual
value of the potential at the realistic minimum, say $V_{\rm
real\;min}$, is important for the CCB analysis since the possible
CCB vacua are dangerous as long as they deeper than $V_{\rm
real\;min}$. From (\ref{Vhiggs}) it is straightforward to get
$V_{\rm real\;min}$
\be
\label{Vreal}
\hspace{-0.1cm}V_{\rm real\;min}
=- \frac{1}{8} (g'^2+g_2^2) (v_2^2-v_1^2)^2 
= - \frac{ \left \{ \left[ ( m_1^2+m_2^2 )^2-4 |m_3|^4  \right]
^{1/2}  - m_1^2+m_2^2  \right \} ^2  } {2 \  (g'^2+g_2^2) }
\ee
Note that this is the result obtained by minimizing
just the tree-level part of (\ref{Vhiggs}).
As explained in sect.2 this procedure is correct if the minimization
is performed at some sensible scale $Q$, which should be of the order
of the most relevant mass entering $\Delta V_1$, see 
eq.(\ref{DeltaV1p}). Since we are dealing here with the 
Higgs-dependent part of the potential, that mass is necessarily
of the order of the largest Higgs-dependent mass, namely the
largest stop mass. From now on we will denote this scale by
$M_S$\footnote{A more precise estimate of $M_S$ was given in
\cite{CCB}, but for our purposes this is accurate enough.}.

Finally, to be considered as realistic, the previous 
minimum must be really a minimum in the {\em whole} field-space. 
This simply implies that all the
scalar squared mass eigenvalues  (squarks and
sleptons) must be positive. Actually, we should go further and
demand that all the not yet
observed particles, i.e. charginos, squarks, etc.,
have masses compatible with the experimental bounds. 
%
%Conservatively enough,
%one can take
%%\cite{particle}
%\bea
%\label{Expb}
%M_{g}&\geq&120\ {\rm GeV}\;,\; \; M_{\chi^{\pm}}\geq 45\ {\rm GeV}
%\nonumber \\
%M_{\chi^o}&\geq&18\ {\rm GeV}\;,\; \; M_{\tilde q}\geq 100\ {\rm GeV}
%\nonumber \\
%M_{\tilde t}&\geq&45\ {\rm GeV}\;,\; \; M_{\tilde l}\geq 45\ {\rm GeV}
%\;\; ,
%\eea
%in an obvious notation. 

\section{Unbounded from below (UFB) constraints}

In this section we analyze the constraints that arise from directions 
in the field-space along which the (tree-level) potential can become 
unbounded from below (UFB). It is in fact possible to give a {\em 
complete} 
clasification of the potentially dangerous UFB directions 
and the corresponding constraints in the MSSM. In order to understand
what are the dangerous directions and the form of the corresponding 
bounds it is useful to notice the following two general properties
about UFB in the MSSM:

\begin{enumerate}

\item[{\bf 1}]
Contrary to what happens to the CCB minima (see sect.1), the
trilinear scalar terms cannot play a
significant role along an UFB direction since for large enough values
of the fields the corresponding quartic (and positive) F--terms
become unavoidably larger.

\item[{\bf 2}]
Since all the physical masses must be positive at $Q=M_S$,
the only negative terms in the (tree-level)
potential that can play a relevant role along an UFB direction
are\footnote{The only possible exception are the stop soft mass terms
$m_{Q_t}^2 |Q_t|^2+m_{t}^2 |t|^2$ since the stop masses are given by
$\sim (m_{Q_t,t}^2 +M_{top}^2 \pm \;{\rm mixing})$, but this possibility
is barely consistent with the present bounds on squark masses.}
\be
\label{ufbneg}
m_2^2 |H_2|^2\;,\;\;\;-2|m_3^2| |H_1||H_2|\;\;\;\;.
\ee
Therefore, any UFB direction must involve, $H_2$ and, {\em perhaps},
$H_1$. Furthermore, since the previous terms are cuadratic, all the
quartic (positive) terms coming from F-- and D--terms must be
vanishing or kept under control along an UFB direction. This means
that, in any case, besides $H_2$ some additional field(s) are required
for that purpose.
In all the instances, the preferred additional fields are 
$H_1$ and/or sleptons since
they normally have smaller soft masses and therefore amount to a less
positive contribution to the potential.

\end{enumerate}
\noindent
Using the previous general properties we can completely clasify the
possible UFB directions in the MSSM. Special attention should be paid
to the UFB--3 bound, which is the strongest one:

\begin{description}

\item[UFB-1]
${}^{}$\\
The first possibility is to play just with $H_1$ and $H_2$. Then,
the relevant terms of the potential are those written
in eq.(\ref{Vhiggs}). Obviously, the only possible UFB direction
corresponds to choose $H_1=H_2$ (up to $O(m_i)$ differences
which are negligible for large enough values of the fields),
so that the quartic D--term is cancelled. Thus, the (tree-level) potential
along the UFB-1 direction is
\be
\label{Vnuevo}
V_{\rm UFB-1}=(m_1^2 + m_2^2 - 2|m_3^2|) |H_2|^2  \ .
\ee
The constraint to be imposed is that, for {\em any} value of $|H_2|<M_X$,
\be
\label{nuevo}
V_{\rm UFB-1}(Q=\hat Q)> V_{\rm real\;min}(Q=M_S) \ ,
\ee
where $V_{\rm real\;min}$ is the value of the realistic minimum,
given by eq.(\ref{Vreal}), and $V_{\rm UFB-1}$ is evaluated at an
appropriate scale $\hat Q$ (see sect.2). 
$\hat Q$ must be of the same order
as the most significant mass along this UFB-1 direction, which is
obviously of order $H_2$. More precisely
$\hat Q \sim {\rm Max}(g_2|H_2|,\ \lambda_{top}|H_2|,$
$M_S)$. Consequently, from (\ref{Vnuevo}) the bound (\ref{nuevo}) is
accurately equivalent to the well-known condition
\be
\label{ufb1}
m_1^2+m_2^2 \ge 2|m_3^2|.
\ee

{}From the previous discussion, it is clear
that the bound (\ref{ufb1}) must be satisfied at any $Q>M_S$ and, 
in particular, at $Q=M_X$.

\item[UFB-2]
${}^{}$\\
If, besides $H_2,H_1$, we consider additional fields in the game, it is 
easy to check by simple inspection (see property 2 above) that the best 
possible choice is a slepton
$L_i$ (along the $\nu_L$ direction), since it has the
lightest mass without contributing to further quartic terms in $V$.
Consequently, from eq.(\ref{V0}), the relevant potential reads
\bea
\label{Vrel}
V&=&m_1^2 |H_1|^2 + m_2^2 |H_2|^2 - 2|m_3^2||H_1||H_2|
+ m_{L_i}^2 |L_i|^2 
\nonumber \\
&+&
\frac{1}{8}(g'^2+g_2^2)( |H_2|^2-|H_1|^2-|L_i|^2 )^2.
\eea
By minimizing $V$ with respect to $H_1, L_i$, it is possible to
write these two fields in terms of $H_2$. This step leads to
non-trivial results provided that $|m_3^2| < \mu^2$,
$|H_2|^2 > 4m_{L_i}^2/ (g'^2+g_2^2)
\left[ 1- \frac{|m_3|^4}{\mu^4} \right]$;
otherwise the optimum value for $L_i$ is $L_i=0$, and we come back to
the direction UFB-1. Then, the potential along the UFB-2 direction
reads\footnote{Eq.(\ref{Vufb2})
relies on the equality $m_1^2-m_{L_i}^2=\mu^2$, which only holds
under the assumption of degenerate soft scalar masses for
$H_1$ and $L_i$ at $M_X$ and in the
approximation of neglecting the bottom and tau Yukawa couplings
in the RGEs. Otherwise, one simply must replace
$\mu^2$ by $m_1^2-m_{L_i}^2$ in eq.(\ref{Vufb2}).} 
\be
\label{Vufb2}
V_{\rm UFB-2}=\left[m_2^2 + m_{L_i}^2 - \frac{|m_3|^4}{\mu^2}\right]
|H_2|^2 - \frac{2m_{L_i}^4}{g'^2+g_2^2}
\;\;.
\ee
 From (\ref{Vufb2}) it might seem that the potential is unbounded from
below unless $m_2^2+m_{L_i}^2-\frac{|m_3|^4}{\mu^2}\geq 0$.
However, strictly, the UFB-2 constraint reads
\be
\label{condufb2}
V_{\rm UFB-2}(Q=\hat Q)> V_{\rm real\;min}(Q=M_S)
\;\;,
\ee
where $V_{\rm real\;min}$ is the value of the realistic minimum,
given by eq.(\ref{Vreal}), and $V_{\rm UFB-2}$ is evaluated at an
appropriate scale $\hat Q$. Again 
$\hat Q \sim {\rm Max}(g_2|H_2|,$
$\lambda_{top}|H_2|,\ M_S)$.

\item[UFB-3]
${}^{}$\\
The only remaining possibility is to take $H_1=0$. Then, the $H_1$
F--term can be cancelled with the help of the VEVs of sleptons
of a particular generation, say $e_{L_j},e_{R_j}$,
without contributing to further quartic terms.
More precisely
\be
\label{trick}
\left|\frac{\partial W}{\partial H_1}\right|^2=
\left|\mu H_2 + \lambda_{e_j} e_{L_j} e_{R_j} \right|^2 = 0
\;\;,
\ee
where $\lambda_{e_j}$ is the corresponding Yukawa coupling.
It is important to note that this trick is not useful if $H_1\neq
0$, as it happens in the UFB--2 direction, since then the 
$e_{L_j},e_{R_j}$ F--terms would
eventually dominate. Now, in order to cancel (or keep under control)
the $SU(2)_L$ and $U(1)_Y$ D--terms we need the VEV of some additional field,
which cannot be $H_1$ for the above mentioned reason. Once again the
optimum choice is a slepton $L_i$ (with $i\neq j$) along 
the $\nu_L$ direction, as in the UFB--2 case.
Denoting $|e_{L_j}| = |e_{R_j}| \equiv$ $|e|$ =
$\sqrt{\frac{|\mu|}{\lambda_{e_j}}|H_2|}$,  the relevant 
potential reads
\bea
\label{Vrel3}
V&=&(m_2^2-\mu^2) |H_2|^2 + (m_{L_j}^2+m_{e_j}^2) |e|^2 + m_{L_i}^2
|L_i|^2 
\nonumber \\
&+&
\frac{1}{8}(g'^2+g_2^2)(|H_2|^2+|e|^2-|L_i|^2)^2.
\eea
Now, the value of $L_i$ can be written, by simple
minimization, in terms of $H_2$, namely 
$|L_i|^2$=$-\frac{4m_{L_i}^2}{g'^2+g_2^2}$+($|H_2|^2$+$|e|^2$).
It turns out that for any value of 
$|H_2|<M_X$ satisfying
\be
\label{SU6}
|H_2| > \sqrt{ \frac{\mu^2}{4\lambda_{e_j}^2}
+ \frac{4m_{L_i}^2}{g'^2+g_2^2}}-\frac{|\mu|}{2\lambda_{e_j}} \ ,
\ee
the value of the potential along the UFB-3 direction is simply
given by
\be
\label{Vufb3}
\hspace{-0.4cm}
V_{\rm UFB-3}=\left[m_2^2 -\mu^2+ m_{L_i}^2 \right]
|H_2|^2 + \frac{|\mu|}{\lambda_{e_j}}\left[m_{L_j}^2 +m_{e_j}^2+ m_{L_i}^2
\right]
|H_2| - \frac{2m_{L_i}^4}{g'^2+g_2^2}
\;\;,
\ee
Otherwise
\bea
\label{Vufb3p}
V_{\rm UFB-3}&=&\left[m_2^2 -\mu^2\right]
|H_2|^2 + \frac{|\mu|}{\lambda_{e_j}}\left[m_{L_j}^2 +m_{e_j}^2\right]
|H_2| 
\nonumber \\
&+&
\frac{1}{8}(g'^2+g_2^2)\left[ |H_2|^2+\frac{|\mu|}{\lambda_{e_j}}
|H_2|  \right]^2.
\eea
Then, the UFB-3 condition reads
\be
\label{SU7}
V_{\rm UFB-3}(Q=\hat Q) > V_{\rm real \; min}(Q=M_S) \ ,
\ee
where $V_{\rm real \; min}$ is given by eq.(\ref{Vreal}),
$\hat Q\sim {\rm Max}(g_2 |e|, \lambda_{top} |H_2|, g_2 |L_i|,$ $M_S)$.

It is interesting to mention that the previous constraint (\ref{SU7})
with the replacements
$e \rightarrow d\;,\hspace{0.1cm}
\lambda_{e_j} \rightarrow \lambda_{d_j}\;,\hspace{0.1cm}
L_j \rightarrow Q_j\;,$
must also be imposed. Now $i$ may be equal to $j$ (the optimum choice is
${d_j}=$ sbottom)
and $\hat Q\sim {\rm Max}\ (\lambda_{top} |H_2|,
\ g_3 |d|,\ \lambda_{u_j} |d|, \ g_2 |L_i|,\ M_S)$.

Anyway, the optimum condition is the one
with the sleptons (note e.g. that the second term
in eqs.(\ref{Vufb3}, \ref{Vufb3p}) is proportional to the 
slepton masses and thus
smaller) and indeed represents, as we will
see in sect.6, the {\it strongest} one of {\it all} the UFB and CCB constraints
in the parameter space of the MSSM.

\end{description}

\section{Charge and color breaking (CCB) constraints}

These constraints arise from the existence of CCB minima in the 
potential deeper than the
realistic minimum. We have already mentioned the ``traditional" CCB
constraint \cite{Frere} of eq.(\ref{frerebound}).
Other particular CCB constraints have been explored in the
literature \cite{Drees,Gunion,Komatsu}.
In this section we will perform a {\em complete} analysis of
the CCB minima, obtaining a set of analytic constraints that represent the
{\em necessary and sufficient} conditions to avoid the dangerous ones. 
As we will see, for certain values of the initial parameters, the CCB 
constraints ``degenerate" into the previously found UFB constraints
since the minima become unbounded from below directions. In this sense, 
the following CCB constraints comprise the UFB bounds of the previous
section, which can be considered as special (but extremely important) 
limits of the former.

In order to gain intuition about CCB, let us enumerate a number of
general properties which are relevant when
one is looking for CCB constraints in the MSSM.
(Formal proofs of the following statements can be found in
ref.\cite{CCB}.)

\begin{enumerate}

\item[{\bf 1}]
The most dangerous, i.e. the deepest, CCB directions in the
MSSM potential involve only one particular trilinear soft
term of one generation (see eq.(\ref{Vsoft})). This can be either of the
leptonic type (i.e. $A_{e_i}\lambda_{e_i}L_{i} H_1 e_{i}$) or the
hadronic type (i.e. $A_{u_i}\lambda_{u_i}Q_{i} H_2 u_{i}$ or
$A_{d_i}\lambda_{d_i} Q_{i} H_1 d_{i}$). Along one of these CCB
directions the remaining trilinear terms are vanishing or negligible.
This is because the presence of a non-vanishing trilinear term in the
potential gives a net negative contribution only in a region of the
field space where the relevant fields are of order $A/\lambda$ with
$\lambda$ and $A$ the corresponding Yukawa coupling
and soft trilinear coefficient;
otherwise either the (positive) mass terms or the (positive)
quartic F--terms associated with these fields dominate the potential.
In consequence two trilinear couplings with different values of
$\lambda$ cannot efficiently ``cooperate" in any region of the field
space to deepen the potential.
Accordingly, to any optimized CCB constraint there corresponds
a unique relevant trilinear coupling, which makes the 
analysis much easier.

\item[{\bf 2}]
If the trilinear term under consideration has a Yukawa coupling
$\lambda^2\ll g^2$, where $g$ represents a generic gauge coupling 
constant, then along the corresponding deepest CCB direction the 
D-term must be vanishing or negligible. This occurs essentially
in all the cases except for the top, and simplifies enormously the 
analysis.

\end{enumerate}

\noindent
{}From the previous properties it can be checked that 
for a given trilinear coupling under consideration
there are {\em two} different relevant directions
to explore. Next, we
illustrate them taking the trilinear coupling of the first
generation, $A_{u}\lambda_{u}Q_u H_2 u_R$, as a guiding example.

\begin{description}

\item[Direction (a)] \hspace{1cm}

It exploits the trick expounded in the direction UFB-3. Namely, if
$H_1=0$, then one can take two $d$-type squarks $d_{L_j}, d_{R_j}$
(or sleptons $e_{L_j}, e_{R_j}$) such that $\lambda_{d_j}\gg\lambda_u$
(or $\lambda_{e_j}\gg\lambda_u$), so that their VEVs cancel the 
$H_1$ F--term, i.e.
\be
\label{H1Fterm}
\left|\frac{\partial W}{\partial H_1}\right|^2=\left|\mu H_2+
\lambda_{d_j}d_{L_j}d_{R_j}\right|^2= 0\;\;.
\ee
Notice that $H_1$ must be very small or vanishing,
otherwise the (positive) $d_{L_j}$ and $d_{R_j}$ F--terms,
$\lambda_{d_j}^2 \left\{ |H_1 d_{R_j}|^2+|d_{L_j} H_1|^2\right\}$,
would clearly dominate the potential (this is also in agreement 
with the property 1 above). 
Since $|d_{L_j}|^2,|d_{R_j}|^2\ll |H_2|^2,|Q_u|^2,
|u_R|^2$, the $d_{L_j}, d_{R_j}$ mass terms are negligible and
the net effect of eq.(\ref{H1Fterm}) is to decrease the $H_2$ 
squared mass from $m_2^2$
to\footnote{Recall that $m_2^2-\mu^2=m_{H_2}^2$, i.e. the $H_2$
soft mass, see sect.3.} $m_2^2-\mu^2$. Furthermore, in addition
to $H_2,Q_u,u_R, d_{L_j},d_{R_j}$, other fields could take
extra non-vanishing VEVs. As in the above-explained UFB-2 direction
(see sect.4) and for similar reasons, it turns out that the optimum
choice is $L_i\neq 0$, with the VEV along the $\nu_L$ direction.
Therefore, along the direction ({\em a})
\subequations{
\be
\label{ccb31}
H_2,Q_u,u_R \neq 0 \;,\;\;
{\rm Possibly}\;\;\;L_i\neq 0 \;\;,
\ee
\be
\label{ccb32}
|d_{L_j}|^2 =|d_{R_j}|^2\ ;\;\; 
d_{L_j}d_{R_j} = -\frac{\mu}{\lambda_{d_j}} H_2
\ee}
\endsubequations
\item[Direction (b)] \hspace{1cm}

If we allow for $H_1\neq 0$, then we cannot play the trick of 
eq.(\ref{H1Fterm}) to cancel the $H_1$ F--term. Therefore, along
this alternative direction
\be
\label{ccb36}
H_2,Q_u,u_R,H_1\neq 0 \;,\;\;
{\rm Possibly}\;\;\;L_i\neq 0 \;\;,
\ee
%
%where $Q_u$ takes the VEV along the $u_L$ direction.

\end{description}

\noindent
Let us now write the potential along the directions ($a$), ($b$).
It is useful for this task to express the
various VEVs in terms of the $H_2$ one, using the following notation
\cite{Gunion}
\bea
\label{alfabeta}
|Q_u|&=&\alpha |H_2|\;,\;\;\;\; |u_R|=\beta |H_2| \;\;,
\nonumber \\
|H_1|&=&\gamma |H_2|\;,\;\;\;\; |L_i|=\gamma_L |H_2| \;\; .
\eea
E.g. the ``traditional" direction, eq.(\ref{frerevevs}), is recovered for the
particular values $\alpha=\beta=1$, $\gamma=\gamma_L=0$.
Now, the basic expression for the scalar potential (see eq.(\ref{V0}))
is
\bea
\label{Valfabeta}
V = \lambda_u^2 F(\alpha,\beta,\gamma,\gamma_L) \alpha^2 \beta^2 |H_2|^4
 -2 \lambda_u \hat A(\gamma) \alpha \beta |H_2|^3
 + {\hat m}^2(\alpha,\beta,\gamma,\gamma_L) |H_2|^2 \;\;,
\eea
where
\bea
\label{FfAm}
F (\alpha,\beta,\gamma,\gamma_L) &=&1+\frac{1}{\alpha^2} +\frac{1}{\beta^2}
+\frac{f(\alpha,\beta,\gamma,\gamma_L)} {\alpha^2\beta^2}\;\;,
\nonumber \\
f(\alpha,\beta,\gamma,\gamma_L)&=& \frac{1}{\lambda_u^2}\left\{
\frac{1}{8} g_2^2 \left(1-\alpha^2-\gamma^2-\gamma_L^2\right)^2\right.
\nonumber \\
\;\;&\ & + \left.
\frac{1}{8}{g'}^2\left(1+\frac{1}{3}\alpha^2-\frac{4}{3}\beta^2
-\gamma^2-\gamma_L^2\right)^2\ +\
\frac{1}{6}g_3^2\left(\alpha^2-\beta^2\right)^2
\right\}\;\;,
\nonumber \\
\hat A(\gamma)&=& |A_u|+|\mu|\gamma\;\;,
\nonumber \\
{\hat m}^2 (\alpha,\beta,\gamma,\gamma_L) &=& m_2^2+m_{Q_u}^2\alpha^2
+m_{u}^2\beta^2 + m_1^2\gamma^2 + m_{L_i}^2\gamma_L^2 - 2|m_3^2|\gamma
\;\; .
\eea
For the ({\em a})--direction eqs.(\ref{Valfabeta},\ref{FfAm})
hold replacing $\gamma=0$, $m_2^2\rightarrow m_2^2-\mu^2$ in
eq.(\ref{FfAm}).
For the ({\em b})--direction, when sign$(A_u)=$ sign$(B)$, it is
not possible to choose the phases of the fields in such a way that
the trilineal scalar coupling ($\propto A_{u} \lambda_{u} Q_u H_2 u_R$),
the cross term in the $H_2$ F--term ($\propto \mu \lambda_{u} Q_u H_1 u_R$)
and the Higgs mixing term ($\propto \mu B H_1 H_2$)
become negative at the same time \cite{CCB}. 
Correspondingly, one (any) of the three terms
$\left\{|A_u|,|\mu|\gamma,-2|m_3^2|\gamma\right\}$ in eq.(\ref{FfAm})
must flip the sign.

Minimizing $V$ with respect to $|H_2|$ for fixed values of
$\alpha,\beta,\gamma,\gamma_L$, we find, besides the $|H_2|=0$ extremal
(all VEVs vanishing), the following CCB solution
\bea
\label{H2min}
|H_2|_{ext}=|H_2(\alpha,\beta,\gamma,\gamma_L)|_{ext}=
\frac{3\hat A}{4\lambda_u\alpha\beta F}
\left\{1\ +\ \sqrt{1-\frac{8\hat m^2 F}{9\hat A^2}}\;\right\}\;.
\eea
\bea
\label{VCCBmin}
V_{\rm CCB\ min}
% = -\lambda_u\alpha^2\beta^2 |H_2|_{ext}^3\left[\lambda_u|H_2|_{ext}F
%-\frac{\hat A}{\alpha\beta}\right]
%\nonumber \\
=-\frac{1}{2}\alpha\beta |H_2|_{ext}^2\left(\hat A \lambda_u |H_2|_{ext}
-\frac{\hat m^2}{\alpha\beta}\right)\;\;.
\eea
Notice that, as was stated above (see property 1), 
the typical VEVs at a CCB minimum are indeed
of order $A/\lambda$. The previous CCB minimum will be
negative\footnote{The mere existence of a CCB minimum is discarded by
demanding ${\hat A}^2 < (8/9)F\hat m^2$, see eq.(\ref{H2min}).}
(and much deeper than the realistic minimum) unless 
\be
\label{CCBgeneral}
{\hat A}^2 \leq F\hat m^2
\ee
This is in fact the most general form of a CCB constraint.

The previous CCB bound takes a more handy form if we realize that
since $\lambda_u^2\ll 1$ (see property 3 above) the D--terms
should vanish. This implies $\alpha^2-\beta^2=0$, $1-\alpha^2-\gamma^2-
\gamma_L^2=0$.
As a consequence $f(\alpha,\beta,\gamma,\gamma_L)$ becomes
vanishing and $F=1+\frac{2}{\alpha^2}$.

Now, we can write the explicite form of the bounds for the directions
({\em a, b}):

\begin{description}

\item[CCB-1]
${}^{}$\\
This bound arises by considering the direction ({\em a}) and thus the
general condition (\ref{CCBgeneral}) takes the form
\bea
\label{GenConda}
\hspace{-0.4cm}
|A_u|^2\ \leq \left(1+\frac{2}{\alpha^2}\right)
\left[ m_2^2-\mu^2+(m_{Q_u}^2+m_{u}^2)\alpha^2
+ m_{L_i}^2\gamma_L^2 \right] \ ,
\eea
where $\alpha^2$ is arbitrary and $\gamma_L^2$ is given
by $\gamma_L^2=1-\alpha^2$. More precisely

\begin{enumerate}
\item If $ m_2^2-\mu^2+m_{L_i}^2>0$ and
$3m_{L_i}^2-(m_{Q_u}^2+m_{u}^2)+2(m_2^2-\mu^2) > 0$,
then the optimized CCB-1 occurs for $\alpha=1$, i.e.
\bea
\label{CCCB1a}
|A_u|^2 \leq  3 \left[ m_2^2-\mu^2+m_{Q_u}^2+m_{u}^2\right]
\eea

\item If $ m_2^2-\mu^2+m_{L_i}^2>0$ and
$3m_{L_i}^2-(m_{Q_u}^2+m_{u}^2)+2(m_2^2-\mu^2) < 0$,
then the optimized CCB-1 bound is
\bea
\label{CCCB1b}
|A_u|^2 \leq \left(1+\frac{2}{\alpha^2}\right)
\left[ m_2^2-\mu^2+(m_{Q_u}^2+m_{u}^2)\alpha^2
+ m_{L_i}^2(1-\alpha^2) \right]
\eea
with $\alpha^2=\sqrt{\frac{2(m_{L_i}^2+m_2^2-\mu^2)}
{m_{Q_u}^2+m_{u}^2-m_{L_i}^2}}.$

\item If $ m_2^2-\mu^2+m_{L_i}^2<0$, then the CCB-1 bound is automatically
violated. In fact the minimization of the
potential in this case gives $\alpha^2 \rightarrow 0$, and we are exactly
led to the UFB-3 direction shown above, which represents the correct
analysis in this instance.

\end{enumerate}

\item[CCB-2]
${}^{}$\\
This bound arises by considering the direction ({\em b}). Then
the general condition (\ref{CCBgeneral}) takes the form
\bea
\label{GenCond}
\left(|A_u|+|\mu|\gamma\right)^2\ &\leq& \left(1+\frac{2}{\alpha^2}\right)
\left[ m_2^2+(m_{Q_u}^2+m_{u}^2)\alpha^2 \right.
\nonumber \\
&+& 
\left. m_1^2\gamma^2 + m_{L_i}^2\gamma_L^2 - 2|m_3^2|\gamma
\right]
\eea
where $\alpha^2, \gamma^2$ are arbitrary and $\gamma_L^2$ is given
by $\gamma_L^2=1-\alpha^2-\gamma^2$.
Rules to handle this bound in an efficient way (i.e. to take the
values of $\alpha^2, \gamma^2$ that make the bound as strong as
possible) can be found in ref.\cite{CCB}.

If sign$(A_u)=$sign$(B)$, the sign of one of the three
terms $\left\{|A_u|,|\mu|\gamma,-2|m_3^2|\gamma\right\}$ in 
(\ref{GenCond}) must be flipped (see comments after eq.(\ref{FfAm})).
Notice that, due to the form of (\ref{GenCond}) flipping the sign of
$|A_u|$ or the sign of $|\mu|\gamma$ leads to the same result.
Therefore, there are only two choices to examine.

\end{description}

\noindent
Concerning the renormalization scale at which the previous CCB-1,
CCB-2 constraints must be evaluated, a sensible choice
is $\hat Q\sim {\rm Max}(M_S, g_3\frac{A_u}{4\lambda_u},$
$\lambda_t\frac{A_u}{4\lambda_u})$, since $H_2\sim
\frac{A_u}{4\lambda_u}$, see eq.(\ref{H2min}).

\vspace{0.2cm}
\noindent
The previous CCB-1, CCB-2 bounds are straightforwardly generalized
to all the couplings with coupling constant $\lambda\ll 1$.
This essentialy includes all the couplings apart from the top.
The generalization to the top Yukawa coupling case is more involved
since $\lambda_{top}=O(1)$, so the D-terms should not be assumed to
vanish anymore. Furthermore, the CCB-1 bounds are not longer applicable
due to the absence of $d$--type squarks such that 
$\lambda_{d_j}\gg\lambda_{top}$. Finally, the associated CCB minima
have in many cases a similar size to the realistic one. So, it
is important to examine explicitely the condition 
$V_{\rm CCB\;min}>V_{\rm real\;min}$.

For more details, the interested reader is referred to ref.\cite{CCB}.

\section{Constraints on the SUSY parameter space}

In sections 4--6 a complete analysis of all the potentially
dangerous unbounded from below (UFB)
and charge and color breaking (CCB)
directions has been carried out. Now, we wish to show explicitely,
through a numerical analysis, the restrictive power of the constraints
on the MSSM parameter space. We will see that this is certainly remarkable.

We will consider the
whole parameter space of the MSSM, $m$, $M$, $A$, $B$, $\mu$, with the only
assumption of universality\footnote{Let us
remark, however, that the constraints
found in previous sections are general and they can also be applied
to the non-universal case.}. Actually, universality
of the soft SUSY-breaking terms at $M_X$ is a
desirable property not only to reduce the number of
independent parameters, but also for phenomenological reasons, particularly
to avoid flavour-changing neutral currents (see, e.g. ref.\cite{Ross}).
As discussed in sect.3, the requirement of correct electroweak breaking
fixes one
of the five independent parameters of the MSSM, say $\mu$, so we are left
with only four parameters ($m$, $M$, $A$, $B$). In order to present the 
results in a clear way we will start by considering the particular case
$m=100$ GeV and $B = A - m$ (i.e. the well--known
minimal SUGRA relation \cite{Weinberg}),
and later we will let $B$ to vary freely.

Fig.1a shows the region excluded by  the ``traditional" CCB
bounds of the type of
eq.(\ref{frerebound}), evaluated at an {\em appropriate} scale
(see sect.2). 
%For a point in the parameter
%space to be excluded we have also demanded that the corresponding
%CCB minimum is deeper than the realistic one (this is especially
%relevant for the bounds coming from the top trilinear term).
Clearly, the ``traditional" bounds, when correctly evaluated, turn out to
be very weak. In fact, only the leptonic (circles) and the $d$--type
(diamonds) terms do restrict, very modestly, the parameter space.
Let us recall here that it has been a common (incorrect) practice
in the literature to evaluate these traditional bounds at all the scales
between $M_X$ and $M_W$, thus obtaining very important (and of course
overestimated) restrictions in the parameter space.
Fig.1b shows the region excluded by the ``improved" CCB constraints
obtained in sect.5. Clearly, 
the excluded region becomes dramatically increased.
Notice also that all the trilinear couplings (except the top one in this
case) give restrictions, producing areas constrained by different
types of bounds simultaneously. The restrictions coming from the UFB
constraints, obtained in sect.4, are shown
in Fig.1c. By far, the most restrictive bound is the UFB--3 one (small
filled squares). Indeed, the UFB--3 constraint is the {\em strongest}
one of {\em all} the UFB and CCB constraints, excluding extensive
areas of the parameter space. This is a most remarkable result.
Finally, in Fig.1d we summarize all the constraints
plotting also the excluded region due to the 
experimental bounds on SUSY
particle masses (filled diamonds). The
finally allowed region (white) is quite small.

How do these results evolve when we vary the values of $m$ and $B$?
The results indicate that the smaller the value of $m$ the more 
restrictive the bounds become (an explanation of this behavior will be
given below). More precisely for $m< 50$ GeV the whole parameter
space becomes forbidden (for any value of the remaining parameters).
So, from UFB and CCB constraints we can conclude
\be
\label{mbound}
m\ge 50\ {\rm GeV} \ .
\ee
Obviously, the limiting case $m=0$ is excluded. This is very relevant 
for no-scale models, since $m=0$ is a typical prediction in that kind of
scenarios. 
Concerning the remaing parameter, $B$, the results indicate that
the larger the value of $B$, the more restrictive the bounds.
In general, for $m\simlt 500$ 
GeV \footnote{Larger values of $m$ start to conflict clearly the 
naturality bounds for electroweak breaking \cite{natu,CC}, so they are 
not realistic.}, $B$ has to satisfy the bound
\be
\label{Bbound}
|B|\simlt 3.5\ m\ .
\ee
Figures illustrating eqs.(\ref{mbound},\ref{Bbound}) can be found
in refs.\cite{CCB,bati}.

\vspace{0.2cm}
\noindent
So far, we have just presented the numerical results in the figs.1a--d
and eqs.(\ref{mbound},\ref{Bbound}) with no attempt to explain the 
physical reasons underlying them. It is, however, very instructive
to examine this question. The first thing to note is that, due to 
their structure, the CCB bounds on $A/m$ 
(see eqs.(\ref{GenConda},\ref{GenCond})) are essentially
$m$--invariant and $B$--invariant. The numerical analysis
confirms this fact \cite{CCB,bati}.
On the other hand, the UFB--3, which is the strongest (CCB
and UFB) bound, becomes more stringent as $m_{H_2}^2=m_2^2-\mu^2$
(i.e. the $H_2$ soft mass) becomes more negative. This is clear from
eqs.(\ref{Vufb3}--\ref{SU7}). The precise value of $m_{H_2}^2$ at low energy depends
on its initial value at $M_X$, i.e. $m$, and on the RG running that, due
to the effect of $\lambda_{top}$, brings $m_{H_2}^2$ to negative values.
Consequently, the smaller $m$ and the larger $\lambda_{top}$, the stronger
the UFB--3 bound becomes. Concerning $m$ this result is certainly well
reflected in eq.(\ref{mbound}). Concerning 
$\lambda_{top}$, since $m_{top}\sim\lambda_{top}\langle H_2\rangle$,
where $\langle H_2\rangle=2M_W^2 \sin\beta / g_2^2$, it is clear that
the smaller $\tan \beta$, the larger $\lambda_{top}$ and therefore
the stronger the UFB--3 bound. But $\tan \beta$ decreases as $B$
increases, thus the form of eq.(\ref{Bbound}).
On the other hand, values of $\tan \beta$ too close to 1 demand a
value of $\lambda_{top}$ at low energy higher than the infrared fixed 
point value, which is impossible to get from the running from high
energies. This fact also contributes to the upper bound on $|B|$,
eq.(\ref{Bbound}). \clearpage

\begin{figure}[h]
%\rule{5cm}{0.2mm}\hfill\rule{5cm}{0.2mm}
%\vskip 2.5cm
%\rule{5cm}{0.2mm}\hfill\rule{5cm}{0.2mm}
\centerline{
\psfig{figure=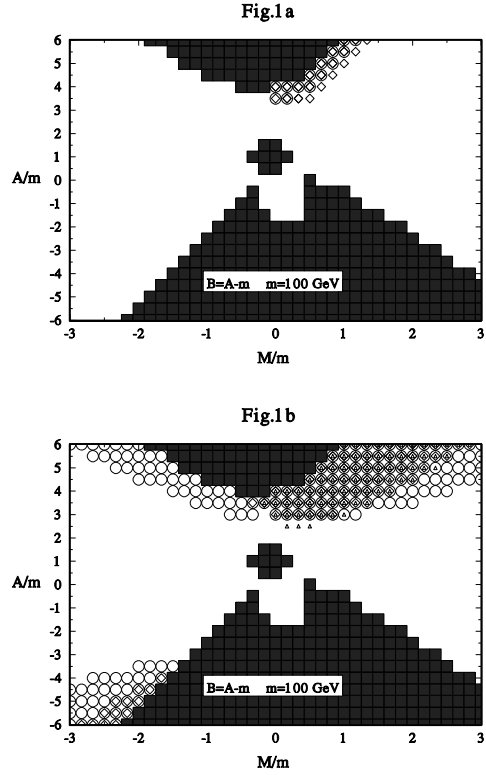,height=10.7cm,angle=180}}
\caption{Excluded regions in the parameter space of the MSSM, 
with $B=A-m$, $m=100$ GeV and
$M^{\rm phys}_{\rm top}=174$ GeV. The darked region is excluded because
there is no solution for $\mu$ capable of producing the correct electroweak
breaking. (a) The circles and diamonds indicate regions excluded by the
``traditional" CCB constraints associated with
the $e$ and $d$-type trilinear terms respectively.
(b) The same as (a) but using the ``improved" CCB
constraints. The triangles correspond to the $u$-type trilinear terms.
(c) The crosses, squares and small filled squares indicate
regions excluded by the UFB-1,2,3 constraints respectively.
d) The previous excluded regions together with the one
arising from the experimental lower bounds on
supersymmetric particle masses (filled diamonds).}
%\caption{bla bla}
\end{figure}

\clearpage

\setcounter{figure}{1}

\begin{figure}[h]
%\rule{5cm}{0.2mm}\hfill\rule{5cm}{0.2mm}
%\vskip 2.5cm
%\rule{5cm}{0.2mm}\hfill\rule{5cm}{0.2mm}
\centerline{
\psfig{figure=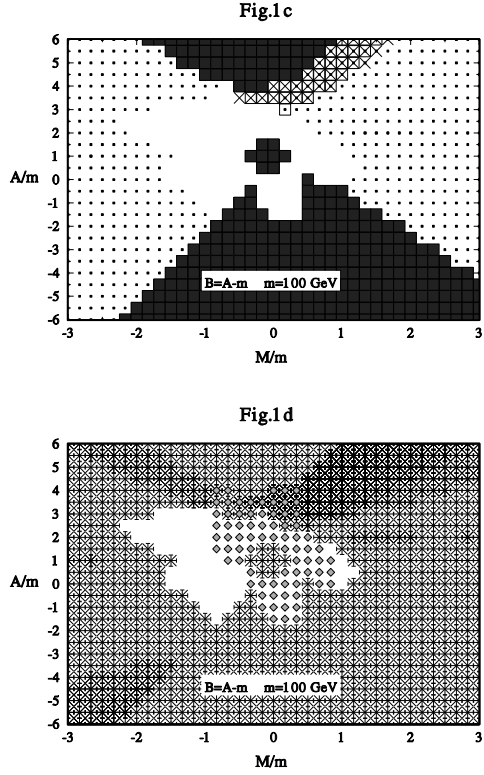,height=10.7cm,angle=180}}
\nonumber
\caption{-- Continued.}
\end{figure}

%
%\vspace{0.3cm}
\noindent
To summarize, the UFB and CCB bounds, specially the UFB-3 bound, 
put important constraints on the 
MSSM parameter space. Contrary to a common believe, the bounds affect
not only the trilinear parameter, $A$, but also the values of the 
universal scalar mass, $m$, the bilinear term parameter, $B$, and
the universal gaugino mass, $M$. This can be noted from the figures 
1a--d and eqs.(\ref{mbound}, \ref{Bbound}). Also, the frequently
used constraint $|A|\le 3m$ is not in general a good approximation. The
actual bounds on $A$ depend on the values of the other SUSY parameters
($m,M,B$).

The application of the UFB and CCB bounds to particular SUSY scenarios
has been considered in some works. It is worth--mentioning that 
the string-inspired dilaton-dominated scenario is completely excluded on
these grounds \cite{dilaton} (as the above-mentioned no-scale scenarios).
The infrared fixed point scenario is also severely constrained
\cite{bati}
(in particular it requires $M<1.1\ m$).

\section{CCB constraints on flavor-mixing couplings}

Supersymmetry has sources of flavor violation which are not
present in the Standard Model \cite{Dim81}. These arise from the
possible
presence of non-diagonal terms in the squark and slepton mass
matrices, coming from the soft-breaking potential (see \footnote{We 
work in a basis for the superfields where the Yukawa coupling matrices
are diagonal.} eq.(\ref{Vsoft}))
\bea
\label{nodiag}
V_{\rm soft}&=&
\left(m_L^2\right)_{ij} \bar L_{L_i}L_{L_j}\ +\
\left(m_{e_R}^2\right)_{ij} \bar{e}_{R_i} e_{R_j}\
\nonumber \\
&+&\
\left(m_Q^2\right)_{ij} \bar Q_{L_i}Q_{L_j}\ +\
\left(m_{u_R}^2\right)_{ij} \bar{u}_{R_i} u_{R_j}\ +\
\left(m_{d_R}^2\right)_{ij} \bar{d}_{R_i} d_{R_j}
\nonumber \\
&+& \left[ A^l_{ij}\bar L_{L_i} H_1 e_{R_j} + A^u_{ij}
\bar Q_{L_i} H_2 u_{R_j}
+ A^d_{ij}Q_{L_i} H_1 d_{R_j}
+ {\rm h.c.} \right] + .. 
\eea
where $i,j=1,2,3$ are generation indices. A usual simplifying
assumption
of the MSSM is that $m^2_{ij}$ is  diagonal and universal and
$A_{ij}$
is proportional to the corresponding Yukawa matrix. 
Actually, we have implicitely used this assumption in all the previous
sections. However, there is
no compelling theoretical argument for these hypotheses
\footnote{This is why, contrary to eq.(\ref{V0}), we have not factorized
the Yukawa couplings, $\lambda$, in the trilinear terms in 
eq.(\ref{nodiag})}.

The size of the off-diagonal entries in $m^2_{ij}$ and $A_{ij}$
is strongly restricted by FCNC experimental data
\cite{Dim81,Choud95,Car95,Gabbi96}. 
Here, we will
focus our attention on the $A_{ij}^{(f)}$ terms; a summary of the
corresponding FCNC bounds is given in the second column of
Table 1 \cite{Choud95,Gabbi96}. The
$\left(\delta^{(f)}_{LR}\right)_{ij}$ parameters used in the table
are
defined as
\bea
\label{notation}
\left(\delta^{(f)}_{LR}\right)_{ij}\equiv
\frac{\left(\Delta
M_{LR}^{2\;\;(f)}\right)_{ij}}{M^{2\;\;(f)}_{av}}\;\;,
\eea
where $f=u,d,l$;
$M^{2\;\;(f)}_{av}$ is the average of the squared sfermion
($\tilde f_L$ and $\tilde f_R$) masses
and $\left(\Delta M_{LR}^{2\;\;(f)}\right)_{ij}$ = $A^{(f)}_{ij}
\langle H_f^0\rangle$, with 
$H_u^0\equiv H_2^0$, $H_{d,l}^0\equiv H_1^0$,
are the off-diagonal entries in the sfermion mass matrices.
It is remarkable that the $A_{ij}^{(f)}$ terms are
also restricted on completely different grounds, namely from the
requirement
of the absence of dangerous charge and color breaking (CCB) minima or
unbounded from below (UFB) directions. These bounds are
in general {\em stronger than the FCNC ones}. Other properties of
these bounds are the following:

\begin{description}

\item[{\em i)}] Some of the bounds, particularly the UFB ones, are
genuine
effects of the non-diagonal $A_{ij}^{(f)}$ structure, i.e. they do
not
have a ``diagonal counterpart''.

\item[{\em ii)}] Contrary to the FCNC bounds, the strength of the
CCB and UFB bounds does not decrease as the scale of supersymmetry
breaking increases.

\end{description}

\noindent
There is no room here to review in detail how these bounds arise,
although the philosophy is similar to that explained in sects.4, 5
(for further details 
%the interested reader is referred to 
see ref.\cite{CasDim}). 
Let us write however the final form of the constraints

\vspace{0.2cm}
{\em {\bf CCB bounds}}
\bea
\label{ccbbound2}
\left|A^{(u)}_{ij}\right|^2 &\leq& \lambda_{u_k}^2\left(m_{u_{L_i}}^2
+
m_{u_{R_j}}^2 + m_2^2\right),
\hspace{1cm}k={\rm Max}\ (i,j)
\nonumber\\
\left|A^{(d)}_{ij}\right|^2 &\leq& \lambda_{d_k}^2\left(m_{d_{L_i}}^2
+
m_{d_{R_j}}^2 + m_1^2\right),
\hspace{1cm}k={\rm Max}\ (i,j)
\nonumber\\
\left|A^{(l)}_{ij}\right|^2 &\leq& \lambda_{e_k}^2\left(m_{e_{L_i}}^2
+
m_{e_{R_j}}^2 + m_1^2\right),
\hspace{1cm}k={\rm Max}\ (i,j)
\eea

\vspace{0.2cm}
{\em {\bf UFB bounds}}

\bea
\label{ufbbound4}
%\hspace{0.2cm}
\left|A^{(u)}_{ij}\right|^2 \leq \lambda_{u_k}^2\left(m_{u_{L_i}}^2 +
m_{u_{R_j}}^2 + m_{e_{L_p}}^2 + m_{e_{R_q}}^2\right),
\hspace{0.8cm}k={\rm Max}\ (i,j),\;\; p\neq q\;.
\nonumber
\eea
\bea
\label{ufbbound3}
\left|A^{(d)}_{ij}\right|^2 \leq \lambda_{d_k}^2\left(m_{d_{L_i}}^2 +
m_{d_{R_j}}^2 + m_{\nu_m}^2\right),
\hspace{1cm}k={\rm Max}\ (i,j)
\nonumber
\eea
\bea
\label{ufbbound2}
\hspace{0.2cm}
\left|A^{(l)}_{ij}\right|^2 \leq \lambda_{e_k}^2\left(m_{e_{L_i}}^2 +
m_{e_{R_j}}^2 + m_{\nu_m}^2\right),
\hspace{0.8cm}k={\rm Max}\ (i,j),\;\; m\neq i,j.
\eea
The CCB bounds must be evaluated at a renormalization scale 
$Q\sim 2A_{ij}^{(f)}/\lambda_{f_k}^2$, while the
UFB bounds must be imposed at any $Q^2
\gg (m/\lambda_{f_k})^2$.
This can be relevant in many instances. For example,
for universal gaugino
and scalar masses ($M_{1/2}$ and $m$ respectively) satisfying
$M_{1/2}\simgt m$, the UFB bounds are more restrictive at $M_X$ than
at low energies (especially the hadronic ones). This trend gets
stronger
as the ratio $M_{1/2}/m$ increases.

\vspace{0.2cm}
\noindent
The previous CCB and UFB bounds can be expressed in terms
of the $\left(\delta^{(f)}_{LR}\right)_{ij}$ parameters defined in
eq.(\ref{notation}) and compared with the corresponding
FCNC bounds. It turns out that the former are almost always stronger.
This is
illustrated in Table 1 for the particular case $M_{av}^{(f)}=500$ GeV.
The only exception is $\left(\delta^{(l)}_{LR}\right)_{12}$, which is
experimentally constrained by the $\mu\rightarrow e,\gamma$ process.
As the scale of supersymmetry breaking increases the FCNC bounds are
easily
satisfied whereas the CCB and UFB bounds continue to strongly
constrain the theory.

Another case in which the FCNC constraints are satisfied
is when approximate ``infrared universality'' emerges from the RG
equations
\cite{gaugino,Choud95,Car95}. Again, the CCB and UFB bounds
continue to impose
strong constraints on such theories. This is because, as argued
before, these bounds have to be evaluated at different
large scales and do not benefit from RG running.

\begin{table}[h]
\caption{FCNC bounds versus  CCB and UFB  bounds on
$(\delta^{(f)}_{LR})_{ij}$ for
$M_{av}^{(f)}=500$ GeV.
The bounds have been obtained from ref.[20]
taking $x=(m_{\rm gaugino}/M_{av}^{(f)})^2=1$.}
\vspace{0.4cm}
\begin{center}
\begin{tabular}{|c|c|c|}\hline
\hspace{0.5cm}&\hspace{0.5cm}&\hspace{0.5cm}\\
\hspace{0.5cm}& FCNC & CCB and UFB \\
%\hspace{0.5cm}&\hspace{0.5cm}&\hspace{0.5cm}\\ 
\hline
%\hspace{0.5cm}&\hspace{0.5cm}&\hspace{0.5cm}\\
$\left(\delta^{(d)}_{LR}\right)_{12}$ & $4.4 \times 10^{-3}$ &
$2.9\times 10^{-4}$\\
%\hspace{0.5cm}&\hspace{0.5cm}&\hspace{0.5cm}\\
$\left(\delta^{(d)}_{LR}\right)_{13}$ &  $3.3 \times 10^{-2}$&
$10^{-2}$\\
%\hspace{0.5cm}&\hspace{0.5cm}&\hspace{0.5cm}\\
$\left(\delta^{(d)}_{LR}\right)_{23}$ &  $1.6\times 10^{-2}$&
$10^{-2}$\\
%\hspace{0.5cm}&\hspace{0.5cm}&\hspace{0.5cm}\\
$\left(\delta^{(u)}_{LR}\right)_{12}$ &  $3.1\times 10^{-2}$&
$2.3\times 10^{-3}$\\
%\hspace{0.5cm}&\hspace{0.5cm}&\hspace{0.5cm}\\
$\left(\delta^{(l)}_{LR}\right)_{12}$ &  $8.5\times 10^{-6}$&
$3.6\times 10^{-4}$\\
%\hspace{0.5cm}&\hspace{0.5cm}&\hspace{0.5cm}\\
$\left(\delta^{(l)}_{LR}\right)_{13}$ &  $5.5\times 10^{-1}$&
$6.1\times 10^{-3}$\\
%\hspace{0.5cm}&\hspace{0.5cm}&\hspace{0.5cm}\\
$\left(\delta^{(l)}_{LR}\right)_{23}$ &  $10^{-1}$&
$6.1\times 10^{-3}$ \\
%\hspace{0.5cm}&\hspace{0.5cm}&\hspace{0.5cm}\\
\hline
\end{tabular}
\end{center}
\end{table}

\section{Cosmological considerations and final comments}

As has been mentioned in sect.1, the CCB and UFB bounds presented 
here are conservative;
they correspond to sufficient, but not necessary, conditions for the
viability of the standard vacuum. It is possible
that we live in a metastable vacuum \cite{Claudson,Kusenko},
whose lifetime is
longer than the age of the universe. This certainly softens
the constraints obtained here.

The first study on CCB-metastability bounds was performed by Claudson 
et al \cite{Claudson}. They showed that only the top-Yukawa 
CCB bounds are dangerous from this point of view. In other words,
among the various CCB minima (see sect.5), the one associated with the
top-Yukawa coupling is the only one to which the realistic minimum has a 
substantial probability to decay during the universe life-time.
The remaining CCB minima, although deeper, present too high barriers
for an efficient tunnelling.
That analysis has been re--done by Kusenko et al. \cite{Kusenko}, taking
into account some subtleties when analyzing the transition 
probabilities. Their results are qualitatively similar to
those of ref.\cite{Claudson}. Quantitatively, they obtain a bound
similar to the CCB-1 bound (see eq.(\ref{CCCB1a})), empirically modified
as $|A_t|^2 \simlt  3 \left[ m_2^2-\mu^2+2.5(m_{Q_t}^2+m_{t}^2)\right]$,
which of course is weaker than the pure stability bound.
On the other hand, the UFB bounds (in particular the UFB-3
bound, which is the strongest of all the CCB and UFB bounds)
have not been analyzed yet from this point of view.

It is important to keep in mind that the metastability bounds
represent necessary but perhaps not sufficient conditions to be safe
(in the same sense that the stability bounds presented in 
sects.4--5 represent sufficient but perhaps not necessary conditions).
The reason is that for the applicability of the metastability bounds,
the universe should be driven by some mechanism into the realistic
(but local and metastable) minimum. This problem has been treated
in several papers \cite{Strumia,Kusenko}. Of course, a 
definite answer (not based in an anthropic principle) requires
the consideration of a particular cosmological scenario in order
to determine the initial values of the relevant fields at early
times. Apparently, the realistic minimum is indeed favoured in many 
cosmological scenarios. Namely, if the initial conditions are dicted by
thermal effects, the universe tends to fall into the realistic
minimum since it is the closest one to the origin.
This can also be the case in some inflationary scenarios. However,
a more systematic analysis of these issues would be welcome.

Finally, from a more philosophic point of view, it is conceptually 
difficult to understand how the cosmological constant is vanishing 
precisely in a local  ``interim'' vacuum (especially from an
inflationary point of view). It is also interesting that 
many of the (tree-level) UFB directions presented here, particularly the
ones associated with flavour violating couplings, are really unbounded 
from below (after radiative corrections) and, if present, make the
theory ill-defined, at least 
until Planckean physics comes to the rescue. These issues, however, 
enter the realm of still unknown pieces of fundamental physics.

\section*{References}

%%%%%%%%%%%%%%%%%%--- References
%%%%%%%%%%%%%%%%%%%%%%%%%%%%%%%%%%%%%%%%%%%%%%%%%%%%%%%
\def\MPL #1 #2 #3 {{\em Mod.~Phys.~Lett.}~{\bf#1}\ (#2) #3 }
\def\NPB #1 #2 #3 {{\em Nucl.~Phys.}~{\bf B#1}\ (#2) #3 }
\def\PLB #1 #2 #3 {{\em Phys.~Lett.}~{\bf B#1}\ (#2) #3 }
\def\PR #1 #2 #3 {{\em Phys.~Rep.}~{\bf#1}\ (#2) #3 }
\def\PRD #1 #2 #3 {{\em Phys.~Rev.}~{\bf D#1}\ (#2) #3 }
\def\PRL #1 #2 #3 {{\em Phys.~Rev.~Lett.}~{\bf#1}\ (#2) #3 }
\def\PTP #1 #2 #3 {{\em Prog.~Theor.~Phys.}~{\bf#1}\ (#2) #3 }
\def\RMP #1 #2 #3 {{\em Rev.~Mod.~Phys.}~{\bf#1}\ (#2) #3 }
\def\ZPC #1 #2 #3 {{\em Z.~Phys.}~{\bf C#1}\ (#2) #3 }

\end{document}